\documentclass[fleqn,3p,sort&compress]{elsarticle}

\usepackage{amsmath}
\usepackage[breaklinks]{hyperref}
\usepackage[latin1]{inputenc}
\usepackage{graphicx}
\usepackage{color}
\usepackage{amsfonts,amsthm,amssymb}
\usepackage{bm,bbm}
\usepackage[OT2,OT1]{fontenc}
\usepackage{slashed}
\newcommand\cyr{%
\renewcommand\rmdefault{wncyr}%
\renewcommand\sfdefault{wncyss}%
\renewcommand\encodingdefault{OT2}%
\normalfont
\selectfont}
\DeclareTextFontCommand{\textcyr}{\cyr}

\def\beq{\begin{equation}}
\def\eeq{\end{equation}}
\newcommand{\be}{\begin{eqnarray}}
\newcommand{\ee}{\end{eqnarray}}

\renewcommand{\texttt}{{}}

\def\bs{\begin{subequations}}
\def\es{\end{subequations}}

\newcommand{\tia}[1]{}

\begin{document}

\begin{frontmatter}

\title{
{\bf Universally Finite Gravitational \& Gauge Theories} } 

\author{Leonardo Modesto} 
\ead{lmodesto@fudan.edu.cn}
\author{Les\l{}aw Rachwa\l{}}
\ead{rachwal@fudan.edu.cn}

\address{
{\small Department of Physics \& Center for Field Theory and Particle Physics,} \\
{\small Fudan University, 200433 Shanghai, China}
}

\date{\small\today}

\begin{abstract} \noindent
It is well known that standard gauge theories are renormalizable in $D=4$ while Einstein gravity 
is renormalizable in $D=2$. This is where the research in the field of two derivatives theories is currently standing.
We hereby present a class of weakly non-local higher derivative 
gravitational and gauge theories universally consistent at quantum level in any spacetime dimension. 
These theories are unitary (ghost-free) and perturbatively  renormalizable. Moreover, we can always find a simple
extension 
of these theories that is super-renormalizable or finite at quantum level in even and odd spacetime dimensions. 
Finally, we propose a super-renormalizable or finite theory for gravity coupled to matter
laying the groundwork for a ``finite standard model of particle physics" and/or a grand unified theory of all
fundamental interactions.

\end{abstract}


\end{frontmatter}
\tableofcontents

\section{Introduction}

One hundred years after the invention of classical general relativity we hereby propose a class of new actions for gravity 
and gauge theories. Theories described by these actions are all 
compatible with quantum mechanics in any spacetime dimension. By the compatibility of these field theories with quantum mechanics we really mean the ability to control infinities, which are ubiquitous at perturbative loop levels. Of course non-renormalizable theories of fundamental interactions may be useful as effective field theories and there is in principle nothing inconsistent in their quantum formulation. However, in this paper we would like to propose theories fully consistent on the quantum level and of fundamental character.
%

We require the following two guiding principles to be common to all fundamental interactions:
``{super-renormalizability or finiteness}" and ``validity of perturbative expansion" in the quantum field theory framework. 
The desired theory satisfies the following postulates:
(i) covariance with respect to spacetime diffeomorphism  or gauge invariance; (ii) weak non-locality (or quasi-polynomiality); (iii) unitarity; (iv) quantum super-renormalizability or finiteness.  
The main difference with quantum perturbative Einstein gravity or standard Yang-Mills theory (or abelian quantum electrodynamics QED\footnote{In this article we will mainly consider non-abelian Yang-Mills theories as the example of gauge theories, but the same analysis can be repeated verbatim also in the case of the abelian electrodynamics.}) lies in the second requirement, which makes possible to achieve 
unitarity and renormalizability at the same time in any spacetime dimension $D$. Next, by choosing a subclass of theories with sufficiently high number of derivatives, we may get even better control over perturbative divergences - we actually may get super-renormalizability. This means that infinities in the perturbative calculus appear only to some finite loop order. Finally, by adding some operators, which are higher in powers of the curvature (or field strength for gauge theories), with specially adjusted coefficients we achieve that perturbative beta functions for all the coupling constants of the theory are consistently set to vanish.
In this way we are able to construct a fully consistent quantum theory for gravitational and gauge interactions 
free of any divergence. 


\subsection{Why Quantum Gravity?}
It is well known that all but one of the fundamental interactions are consistent with quantum mechanics. In Nature we observe these interactions in the quantum domain. Only electromagnetism and gravitation have also impact on classical macroscopic world. Actually gravitation as described by general relativity was observed and tested only on the classical level.
In this short paragraph we would like to emphasize theoretical reasons why gravity has to be quantum as well. We give one main reason out of many that can be met in literature.
We know that astrophysical compact objects release massless spin two gravitational waves compatible with classical
linearized Einstein gravity. Astrophysical binary systems are indeed antennas for 
the generation of gravitational waves as well as conductors are antennas for electromagnetic waves.
Therefore, the following theory for a massless particle of spin two 
is realized in Nature and has been already tested by the indirect observation supporting the existence of gravitational waves, 
\be
\Box h_{\mu \nu} = 8 \pi G_N T_{\mu\nu} \, .
\label{linEinth}
\ee
Here $T_{\mu\nu}$ is the energy-momentum tensor of matter.  About the source part of the above equation, we well know that it is of quantum origin, namely it arises from classical expectation values of fluctuations of the quantum matter fields. Indeed all the physics of the microstructure confirms the quantum nature of matter. 
This is our undeniable observational starting point. The theory (\ref{linEinth}) is gauge covariant with respect to linearized gauge transformations of the spin two field. Since such gauge changes can only be the infinitesimal version of general coordinate transformations we also know what are all possible interaction vertices compatible with the full symmetry. 
At this stage the outcome is an action made of an infinite number of gravitational curvature operators 
compatible with diffeomorphism invariance. In technical terms a Lagrangian describing consistently the relativistic gravitational field must be constructed using only generally covariant tensors, covariant derivatives and the metric tensor. Moreover, it must be integrated with the proper measure on the differential spacetime manifold to produce an invariant action functional for the classical gravitational field. 

Now the crucial point is that 
a gravitational field consistently interacting with quantum matter must be quantum itself. We know from basic principles of quantum mechanics that two systems may interact via classical interactions only if both are classical. If the internal dynamics in one of the systems is quantum, then also the mediating interactions must be quantum and in the result also the internal dynamics of the second system must be quantum. Otherwise we would run into very serious conflict with quantum mechanics. A physical example of such astrophysical systems we presented above. Another example of such systems 
is given by neutrons falling in the Earth gravitational field and for which 
quantum interference experiments were already performed. In conclusions we derive here that gravitational interactions with matter must be quantized and for consistency also self-interactions of the gravitational field must be quantized.
If the reader agrees with the content of this paragraph, then the quantization of gravity is unavoidable. A fully consistent theory of gravitational interactions necessarily must be of quantum nature.

\subsection{What does Renormalizability mean?}
We previously mentioned renormalizability or in its stronger version super-renormalizability or even finiteness, 
as one of the ways by which we can have some control over divergences in quantum field theory.
We must emphasize
here that the renormalizability property of a theory is a purely theoretical concept. Experimentalists never can prove that this or that sector of the quantum world is described by a renormalizable theory. It is an abstract concept about our models, which describe physical reality, and as a such never admits direct experimental verification. Moreover, in the high energy experiments the infinities can never be measured, so we do not know empirically how many operators we have to renormalize in a consistent theoretical description. Obviously from the practical point of view of computations done in the quantum field theory (QFT) framework, renormalizability is a very nice property because it constrains strongly the number of counterterms needed for the absorption of perturbative divergences at any loop order. It is a common belief that only renormalizable theories are predictive because 
only for them we need to fix by experiments  a finite number of parameters (to set renormalization conditions and hence renormalize these couplings.)
However, this is not true because empirically we can not measure with infinite precision 
all the possible couplings to really prove that only a finite number of operators defines the theory. This is even more true in the effective field theory framework, where the existence of higher order additional operators is not excluded {\it a priori}. They may be added to the effective theory description, but their effects are highly suppressed at low energies. In the effective theories framework we conclude that theories with non-gravitational interactions may look renormalizable at low energies. However, if they are non-renormalizable at higher energy, then typically they require an ultraviolet (UV) completion and can not be viewed as fundamental. With gravity the situation is more complicated as we will explain in the next section. Perturbative renormalizability is a very natural requirement for the UV-completion of the fundamental theories of gravitational and gauge interactions. 

Our notion of perturbative renormalizability\footnote{There exists also a notion of non-perturbative renormalizability also known as asymptotic safety \cite{ASref}, but we will not discuss it here.} reads as follows. We start with a classical action made of local and weakly 
non-local operators. Then we say that the theory is renormalizable if we can absorb all the quantum divergences in a finite
number of {\em local} operators already present in the classical action. We compare this notion with ``{extended}" renormalizability
in the modern sense stating that all the divergences can be cancelled by renormalization
of the infinite number of terms in the bare action \cite{Gomis, AnselmiQG}. However, here we renormalize up to a finite number of running coupling constants. Or even better, for the case of a finite theory we do not make any infinite renormalization at all. We want to remark here that the standard criterion for renormalizability is not valid for higher derivative theories. Naively we would say that theories are renormalizable if they do not possess dimensionful parameters with negative mass dimension. But this implicitly assumes that the coupling in front of the kinetic term for the fluctuations is dimensionless, which is always the case only for standard two-derivative theories of scalar and fermion fields with canonical kinetic terms. Another hidden assumption is that the dimensions of all  fluctuations are the same. In 
multiplicative renormalizability 
if the number of derivatives in the interactions exceeds the highest number of derivatives in the kinetic term then the
theory is non-renormalizable. 
Moreover, we should take proper care of the energy dimension of the fluctuations. If the fluctuations have all the same energy dimensions, then this analysis goes verbatim also for energy dimensions of couplings in front of the corresponding terms. The couplings in front of the highest derivative kinetic terms may not be dimensionless. For renormalizability of these kinetic terms it is 
only demanded that their frontal coefficients are not with positive energy dimension. They must be with negative or zero energy dimensions because only in this case we can absorb all perturbative divergences appearing at loop levels. (In $D$ spacetime dimensions the divergences in momentum space in a renormalizable theory can be up to the order $k^D$, hence in the kinetic term we need up to $D-2x$ derivatives, where $x$ is the energy dimension of fluctuations). Within this generalized notion of renormalizable theories much depends on the form of the kinetic term and the inclusion of other fluctuating fields may slightly complicate the analysis. Generally for a renormalizable theory in any dimensions $D$ we must include in the bare action all constructible operators allowed by the symmetry 
and having non-negative energy dimension of frontal coefficients. 

Here we want to point out only one additional fact. In this light the statement that an operator is renormalizable or not is very dependent on the form of the kinetic term in the theory and as a such does not have an absolute meaning. Some high derivative operators may be renormalizable or even necessary for renormalizability in some higher derivative theories. In other theories they may spoil this property once they are introduced. Hence the simple statement that operators with many derivatives (and so with negative energy dimension of frontal couplings) are non-renormalizable is true only in low-derivative theories. It is possible to speak about the character of an operator, only when it is added to the already existing theory. However, we see that 
it would be more appropriate and correct to say something only about whether the full theory is renormalizable or not. Yes, we want to associate the notion of renormalizability to a given theory, but not to a particular operator in the theory. We understand that such analyses of renormalizability were first performed for the case of all 2-derivative theories because they are the simplest which give propagation of modes, and because they contain the minimal number of derivatives for the bosonic fields. However, for higher derivative theories these analyses must be refined. We provide such discussions for gravity and gauge theories in any dimension $D$ in the next section.

\subsection{Think back gauge and gravitational theories: accidentally renormalizable theories}

In this section we would like to point out what stopped scientists in the last $30$ - $40$ years to find a new theory of gravity fully compatible with quantum mechanics and remaining within the ``{quantum field theory framework}". 
All but one fundamental interactions in Nature are described by the Yang-Mills action, proportional to 
$
\int\! d^Dx\,g^{-2}{\rm tr}F^2$,
 and it turns out 
that this theory is compatible with the standard quantization paradigm
(in the case of the standard model of particle physics (SM) 
the presence of massive matter makes things less trivial.)
However, it is only in $D=4$ spacetime dimensions, that the Yang-Mills theories (and also the SM) are consistent 
because perturbatively renormalizable. These theories are known to be with only two derivatives and hence the problem with unitarity does not arise here. Moreover, their classical dynamics is completely under control without any instabilities. It is easy to understand why $D=4$ is an accident here. The argument is of dimensional character. The gauge potential $A_\mu$ has always energy dimension of mass irrespectively of the dimensionality $D$ of the spacetime. This is because $A_\mu$ appears in the expression for covariant derivative hand in hand with ordinary partial derivative, whose dimension is equal to mass. 
Here it is also crucial the fact 
that the charge (YM coupling $g$) is dimensionless as well as the parameter of gauge transformation, which appears in the exponent of the finite gauge changes. Requiring the theory of gauge potentials to be with two derivatives only, we find that exactly in $D=4$ its action is without any dimensionfull parameter. Or in other words the YM coupling is in such circumstances dimensionless and hence the theory is perturbatively renormalizable. The propagator for the quantum modes is unitary because it falls off in UV like $1/k^2$, where $k$ is the momentum of the mode.
On the other hand standard gravity is described by Einstein-Hilbert (E-H) action proportional to 
\be
\int\! d^Dx\, 2 \, 
\kappa_D^{-2}\sqrt{|g|} R \, ,
\ee 
which is consistent with quantum mechanics only in $D=2$, but non-renormalizable in $D>2$. This theory is, as it is well known, a theory with two derivatives on the metric. Again here two spacetime dimensions are accidental because only in $D=2$ the E-H dynamics is renormalizable. However, in any dimension this theory is unitary and only the two degrees of freedom of the massless transverse graviton (with two helicities: $+2$ and $-2$) propagate. On the classical level this theory around flat Minkowski background is known to be perfectly stable. We understand the speciality of two dimensions from similar arguments as before. Here as gauge potential we choose the metric (or strictly its fluctuations $h_{\mu\nu}$) that always can be taken as dimensionless no matter in which spacetime and no matter in which dimension $D$ we are working. This is because the metric must measure the distances using already dimensionfull coordinates and hence does not carry any energy dimension. 
The metric does not appear in the definition of the covariant derivatives on the curved manifold, it is its first derivative in the form of Levi-Civita connection to appear there. If we want to construct  a theory of the metric with two derivatives then the Lagrangian density will have energy dimension two. This means that only in $D=2$ spacetime dimensions the action for gravity will be without any dimensionful constant and hence renormalizable. This is in strong contrast to the previous case of gauge theories. The main difference lays in the fact that the dimensionality of gauge potentials in these two cases are different. In higher dimensions the 
quantum theory based on E-H classical action is non-renormalizable due to energy dimension of the gravitational Newton's constant, which is $[G_N\sim\kappa_D^{2}]=M^{2-D}$. Here about the breakdown of renormalizability we can easily argue from the simple Dyson criterion. In our theory we have the presence of the dimensionfull coupling $\kappa_D$. The situation with the behaviour of the graviton propagator is the same as in other two-derivative theories. We can phrase the lack of renormalizability on the loop level as the insufficient 
suppression of the propagation of modes via powers of momentum in the propagator. 
Simply the graviton propagator falls off too slowly in UV as to be described as the effect of a renormalizable dynamics of dimensionless fluctuations. For a dimensionless field fluctuation $h_{\mu\nu}$, in $D$ spacetime dimensions, the propagator should decay at least like $k^D$ in UV regime to comply with perturbative renormalizability. If we are above the critical dimension 
$D=2$ for gravity, then such behaviour of the propagator leads to serious problems with unitarity. We may encounter such problematic situations only in higher derivative theories. 

The problem of unitarity in higher derivative theories also can be easily understood. Already at 
the classical level here one meets the problems of Ostrogradsky instabilities. On the quantum level instead the leading behaviour of the propagator at high energy is usually analyzed. We recall that standard particle interpretation of QFT excitations is valid only for simple fractions with the denominator $k^2 + m^2$, which result from the decomposition of the propagator viewed as a rational function in momentum $k$. For multipole poles in the $k^2$ variable the residue vanishes and hence we do not see the effect of propagation of these modes in the perturbative Feynman propagator. Even if we restrict to the case of single poles in $k^2$, then we can meet two cases, which make our theory non-unitary. In the first case the sign of the residue is negative and again here we lose the interpretation of particle modes propagation. Now their kinetic term is with the wrong sign, it is not bounded from below and the particles are of phantom character. Another pathology is when the mass parameter in the simple pole decomposition is imaginary and then the fraction is of the type $1/(k^2-m^2)$. In this case we speak about tachyons. All these two cases of phantoms and tachyons are examples of very unwanted additional unphysical modes in our theory (so called bad ghosts or poltergeists and tachyons.) It is easy to see that we generate them unavoidably if we want the behavior of the propagator like in higher derivative theories. If the decomposition of the propagator into simple fractions is valid at all energies (this assumes the polynomial character of the higher derivative theory), and if in the UV the propagator decays like $1/k^D$, where $D>2$, then necessarily in this decomposition we will find a simple fraction with negative residue (so at least one phantom mode). Simple fraction with single pole in $k^2$ variable has always decay like $1/k^2$ in UV. Since $D>2$, then according to our requirements the leading UV behaviour of the full propagator must be reduced from $1/k^2$ (from each simple fraction)
to $1/k^D$. This signifies that between simple fractions must happen a cancellation of the highest powers in momentum. This is only possible, if at least one of the residue is negative.

The previous comparison shows the following parallelism between gauge and gravitational theories:
\be
 {\int \!d^4 x \, g^{-2} \, {\rm tr} F^2 \,\,\, {\rm in} \,\,\,\, D=4} \,\,\, \, \mbox{like}  \,\,\,  \int \!d^2x \, 2 \, \kappa_2^{-2} \sqrt{|g|} R 
 \,\,\,\,\, {\rm in} \,\,\, D=2.
\ee
We know how to construct a renormalizable gauge theory in any dimension $D$ (different from 4). We must extend Yang-Mills action by adding higher derivative operators that have the impact on the kinetic term for the gauge potentials. This tells us that we should use higher derivative gauge theories in higher dimension than $D=4$. In the same spirit we can construct renormalizable theories of quantum gravity in any dimension higher than the critical $D=2$ (for example in the relevant case of four spacetime dimensions we should include two new terms to the Einstein-Hilbert action, namely $R^2$ and $R_{\mu\nu}^2$ \cite{Stelle}.)
Let us ask now the following question. 
What about a gauge theory, which is not only renormalizable, but also unitary in any dimension?
If we want to construct a theory renormalizable and unitary for $D>4$ without to introduce other (unwanted) degrees of 
freedom it is unavoidable to introduce a non-local action principle. In this way we can avoid usage of polynomial higher derivative theories that in turn have problems with unitarity as elucidated above. Actually, the non-local theory can be viewed as a slight generalization of the polynomial local higher derivative theory, when we abandon locality, because the degree of derivatives is formally infinite there. In this sense we also move from polynomials in momenta to entire functions of momenta, which however have analytic truncation in the former, when the theory is analyzed in Fourier space. 
Is this telling us, that there is something special in $D=4$? We do not think so because $D=4$ is special only for gauge theory ($D=4$ is the dimension in which YM theory is clasically conformally invariant), but not for gravity. On the other hand $D=2$ is special for gravity (which is conformally invariant in two dimensions only), but not for gauge theory.
Actually, it is impossible
to write a gauge-invariant, local, unitary and perturbatively renormalizable theory of gravity mediated by a massless spin two particle in $D=4$. This would be possible only in the critical dimension $D=2$. 

If we want a unitary and renormalizable theory of quantum gravity above $D=2$, then in full similarity with the case of gauge interactions, we must resign from locality.
Therefore, there is nothing special in describing the dynamics of gravity using a non-local action principle, but 
this is actually the most simple, conservative and fully consistent approach beyond Einstein dynamics. As we see the procedure is exactly the same for gravity and gauge theory, therefore we speak about the necessity of universality 
 for having weakly non-local theories in any dimension to describe field theory models consistent with quantum mechanics.
A further comment regarding the non-polynomial or non-local nature of the action is needed. People are usually skeptic about non-locality in general, overlooking that all the known interactions are characterized by already non-local effective actions \cite{Maggiore} starting at one-loop order. What we did here is just to start out with a non-polynomiality in the classical action. Within the quantum field theory framework, if we want to preserve Lorentz or diffeomorphism invariance and unitarity, while at the same time to have a renormalizable theory of gravity, then we are forced to introduce at least one entire function in the action functional.

Moreover, if we define a unitary and super-renormalizable/finite non-local theory of gravity in $D>4$, then the Kaluza-Klein compactification gives as back a non-local and super-renormalizable Yang-Mills 
theory in one dimension less. Therefore, weakly non-local gauge theories naturally emerge from gravity after compactification. Finally, many people believe in a gauge/gravity correspondence or AdS/CFT duality as a fundamental requirement that any consistent quantum gravitational theory has to satisfy. Therefore, a well defined gravitational theory in the ultraviolet regime must be dual of the above weakly non-local gauge theory. In this way we put much of the importance on duality and the uniform theoretical framework needed for the description of gravity and gauge theory.
In the next part we will present weakly non-local theories of gravitational and gauge interactions treated on the same footing. They possess the desired properties of unitarity and renormalizability. Moreover, within the common framework for these theories, we will be able to find the conditions necessary for super-renormalizability and eventually we will show to the reader our proposals for finite theories of coupled system consisting of gravity, gauge interactions, and matter.

\vspace{-0.2cm}

\section{ 
Gravitational and Gauge theories}
In the previous introduction we have seen that the Yang-Mills theory is special in $D=4$, while Einstein-Hilbert gravity is special 
in $D=2$. But what about the construction of a theory consistent in any dimension? It must be a theory governed by a non-local Lagrangian as we extensively explained in the previous section. A consistent gauge invariant theory for spin one or spin two massless particles 
regardless of the spacetime dimension fits in the following general class of theories 
\cite{Deser, Odintsov, Moffat3, corni1}, 
\be
&&
\boxed{
\mathcal{L}_{\rm YM} = - \frac{1}{4 g_{\rm YM}^2} \left[  {\rm tr} \,  {F} e^{H(-{\cal D}^2_{\Lambda})} {F} 
+ \mathcal{V}_{\rm YM} \right] }
 \,\,\,\,\,\, (\rm GAUGE-THEORY) \, , \label{gauge}
\ee
\be
&&\boxed{ \mathcal{L}_{\rm gr} = -  2 \kappa_{D}^{-2} \, \sqrt{|g|} 
\left[ R 
- \frac{1}{2} \, R 
\, \frac{e^{H(-\Box_{\Lambda})} -1}{\Box}
 R 
 +  R_{\mu\nu} 
\, \frac{e^{H(-\Box_{\Lambda})}-1}{\Box} 
 R^{\mu\nu} 
+ \mathcal{V}_{\rm gr}
\right] 
}
 \, \,\,\,\,\,\,  ({\rm GRAVITY}) \, .
\label{gravity}
\ee
The theories above 
consist of a kinetic weakly non-local operator and a local curvature potentials $ \mathcal{V}$ respectively for the case of gauge and gravitational theory. It is not necessary to include weak non-locality in the potential $\cal V$. 
However, this potential is crucial to achieve finiteness of the theory as we will show later. 
 In (\ref{gauge}) the Lorentz indices and tensorial structures have been neglected.
The notation for the gauge theory (initially put on the flat spacetime) is as follows: we use the gauge covariant box operator defined via ${\cal D}^2={\cal D}_\mu{\cal D}^\mu$, where ${\cal D}_\mu$ is a gauge covariant derivative (in the adjoint representation) acting on gauge covariant field strength $F_{\rho\sigma}$. We will not write in this whole article the gauge group indices that appear on gauge field strengths and on covariant derivatives.
Moreover, for the case of gravity theory 
$\Box = g^{\mu\nu} \nabla_{\mu} \nabla_{\nu}$ is the diffeomorphism covariant box operator\footnote{{\em Definitions ---} 
The metric tensor $g_{\mu \nu}$ has 
signature $(- + \dots +)$ and the curvature tensors are defined as follows: 
$R^{\mu}_{\nu \rho \sigma} = - \partial_{\sigma} \Gamma^{\mu}_{\nu \rho} + \dots $, 
$R_{\mu \nu} = R^{\rho}_{\mu  \rho \nu}$,  
$R = g^{\mu \nu} R_{\mu \nu}$. With symbol ${\cal R}$ we generally denote one of the above curvature tensors.}. For both theories in (\ref{gauge}) and (\ref{gravity}) we employ the following definitions, ${\cal D}^2_{\Lambda} \equiv  {\cal D}^2/\Lambda^2$ and $\Box_{\Lambda} \equiv  \Box/\Lambda^2$ , where $\Lambda$ is an invariant mass scale in our fundamental theory. 
The entire function 
$H(z)$ in (\ref{gauge}) and (\ref{gravity}) will be shortly defined. For it we need the definition of $\rm{N}$ that is the following function of the spacetime dimension $D$: $2 \mathrm{N} + 4 = D$ for even dimensions and $2{\rm N}+3=D$ for odd dimensions.

Finally, the entire function $V^{-1}(z) \equiv \exp H(z)$ ($z \equiv  -{\cal D}^2_\Lambda\,\,{\rm or}\,\,-\Box_{\Lambda}$ respectively)
satisfies the following general conditions \cite{Tombo}:
(i) $V^{-1}(z)$ is real and positive on the real axis and it has no zeros on the 
whole complex plane $|z| < + \infty$. This requirement implies, that there are no 
gauge-invariant poles other than for the transverse and massless gluons and gravitons respectively.
%
$ $(ii) $|V^{-1}(z)|$ has the same asymptotic behavior along the real axis at $\pm \infty$; 
(iii) There exists $\Theta>0$ and $\Theta<\pi/2$, such that asymptotically
\be
|V^{-1}(z)| \rightarrow | z |^{\gamma + \mathrm{N}+1},\,\, {\rm when }\,\, |z|\rightarrow + \infty\,\,\,\, {\rm with}
 \,\,\,\,\,
\gamma\geqslant D_{\rm even}/2 \,  ,
\label{tombocond}
\ee 
for complex values of $z$ in the conical regions $C$ defined by: 
$C = \{ z \, | \,\, - \Theta < {\rm arg} z < + \Theta \, ,  
\,\,  \pi - \Theta < {\rm arg} z < \pi + \Theta\}.$
The last condition is necessary to achieve the maximum convergence of the theory in
the UV regime.  
The necessary asymptotic behavior is imposed not only on the real axis, but also on the conical regions, that surround it.  
In an Euclidean spacetime, the condition (ii) is not strictly necessary if (iii) applies\footnote{A string inspired theory with exponential form factor has been extensively studied in \cite{MazumdarSbagliato}.}.

An explicit example of weakly non-local form-factor $e^{H(z)}$, that has the properties (i)-(iii) can be easily constructed following \cite{Tombo}, 
\be
\hspace{-0.8cm}
 e^{H(z)}
= e^{\frac{1}{2} \left[ \Gamma \left(0, p(z)^2 \right)+\gamma_E  + \log \left( p(z)^2 \right) \right] }=
e^{\frac{\gamma_E}{2}} \,
\sqrt{ p(z)^2} 
\label{Vlimit1} 
\left\{ 
1+ \left[ \frac{e^{-p(z)^2}}{2 p(z)^2} \left(  1 
+ O \left(   \frac{1}{p(z)^2} \! \right)   \! \right) + O \left(e^{-2p(z)^2} \right)  \right] \right\}\label{VlimitB}
\footnote{
In (\ref{VlimitB}), 
 $\gamma_E \approx 0.577216$ is the Euler-Mascheroni constant and 
$
\Gamma(0,x) = \int_x^{+ \infty}  d t \, e^{-t} /t 
$ 
is the incomplete gamma function with its first argument vanishing.} \!\! .
\ee
The  polynomial $p(z)$ of degree $\gamma +\mathrm{N}+1$ is such that $p(0)=0$, which gives the correct low energy limit of our theory (coinciding with the standard two derivative theories of Yang-Mills and Einstein gravity.) In this case the angle $\Theta$ defining the 
 cone $C$ turns out to be $ \pi/(4 (\gamma + \mathrm{N} + 1))$. 
For renormalizability we must require that the integer $\gamma$ is non-negative. We also observe
that the minimal choice $\gamma=0$ restores, in UV,  minimal higher derivative theory with precisely four derivatives. 
Though with this construction we are unable to come back to standard two-derivatives Yang-Mills action in UV. Since in the UV we have still polynomial behavior, but all the problems with unitarity will be sent away, this step gives a revival to all polynomial higher derivative theories regarding questions about divergences and beta functions in particular.
 
The most general local potential for gravity is made of the following three sets of operators (by potential we mean here terms that do not have any influence on the propagator in opposition to kinetic terms, and therefore 
they are higher than quadratic in general gravitational curvature $\cal R$): 
\be
&& \hspace{-1.4cm}
 \mathcal{V}_{\rm gr} = 
 \sum_{j=3}^{{\rm N}+2} \sum_{k=3}^{j} \sum_i c_{k,i}^{(j)} \left( \nabla^{2(j-k)} {\cal R}^k \right)_i \!
 +
  \!\!\!
 \sum_{j={\rm N}+3}^{\gamma+{\rm N}+1} \sum_{k=3}^{j} \sum_i d_{k,i}^{\,(j)} \left(\nabla^{2(j-k)} {\cal R}^k \right)_i
 + \!\!\!
 \underbrace{\sum_{k=3}^{\gamma +{\rm N}+2} \!\! \sum_i s_{k,i}  \left(  \nabla^{2 (\gamma + {\rm N}+2 -k )} \, {\cal R}^k \right)_i}_{\rm killers} \! .  
 \nonumber\\
 &&
 \label{K0}
 \ee 
 The operators in the third set are called killers because they are crucial in making the theory finite in any dimension. The same kind of potential can be written for gauge bosons, we only need to substitute the gravitational curvatures $\cal R$ with the gauge field strengths $F$ and the covariant derivatives 
 $\nabla$ with the gauge covariant ones $\cal D$. For the case of non abelian gauge groups we have many possible ways to take traces over group indices and this freedom is also denoted in the gravitational case by multiple contraction of indices, represented above by an index $i$ in (\ref{K0}). For more details about the notation in the gravitational case we refer to \cite{modestoLeslaw}. Actually for the gauge theory we must end up with a potential ${\cal V}_{\rm YM}$ completely blind in group indices and to achieve this goal we may construct various products of traces of products of field strengths and covariant derivatives (with at least two factors under each trace.)
 
The theories described by the actions in (\ref{gauge}) and (\ref{gravity}) are unitary and perturbatively renormalizable at quantum level in any dimension. They propagate only quantum modes of respectively massless spin one and massless spin two with highest absolute value of helicity. Of course this is true in any dimension $D$. The behaviour of the propagator for all modes is enhanced in the UV regime to accord with renormalizability properties. Hence we conclude that these theories are consistent with quantum mechanics and the control over perturbative divergences is fully recovered. Moreover at classical level many evidences endorse that we are dealing with ``{\em gravitational and gauge theories possessing singularity-free exact solutions}". Discussions for the interesting case of 
cosmology, black holes, and gravitational potential 
can be found in
 \cite{BiswasSiegel,ModestoMoffatNico,BambiMalaModesto2, BambiMalaModesto,calcagnimodesto, koshe1, ModestoGhosts}. 

\subsection{Propagator \& Unitarity}
\label{gravitonpropagator}
Splitting the spacetime metric or the gauge field into a background field plus a fluctuation
(for gravity we expand the metric around the Minkowki spacetime, while for the gauge theory we expand the gauge field around 
a background without any chromo-electric or chromo-magnetic field, namely flat gauge connection),
fixing the gauge freedom and computing the quadratic action for the fluctuations, we can invert the quadratic operator to get finally the two point function  \cite{HigherDG}. This quantity, also known as the propagator in the Fourier space, and up to gauge dependent components, reads
\be
&& 
\mathcal{O}^{-1} \!=\!
 \frac{V(  k^2/\Lambda^2 )  } {k^2}  \, \times \,({\rm TS})\,, \nonumber \\
&& {\rm TS}_g = \left( P^{(2)} - \frac{P^{(0)}}{D-2 }  \right) \, , \,\,\,\,
 {\rm TS}_{\rm YM}=\left( \eta_{\mu\nu} - \frac{k_\mu k_\nu}{k^2} \right) \, ,
 \label{propagator}
\ee
where the symbol ``TS" stands for tensorial structure suitable for the given fluctuation fields. The tensorial structure for gravity is 
${\rm TS}_g$. 
The Lorentz indices for the operator $\mathcal{O}^{-1}$ and the projectors   $\{P^{(0)},P^{(2)}
\}$ from \cite{HigherDG, VN} have been omitted. 
For the gauge theory the tensorial structure above amounts to the standard transverse massless 
projector ${\rm TS}_{\rm YM}$.
 
The tensorial structure in (\ref{propagator}) is the same of Einstein gravity or Yang-Mills theory (both standard with two derivatives), but we see the presence of a new element - the multiplicative 
 form-factor $V(z)$. If the function $V^{-1}(z)$ does not have any zeros on the whole complex plane, then in consequence $V(z)$ does not have any poles and hence the structure of poles is the same as in original two-derivative theory. This means that in the spectrum we have exactly the same modes as in two-derivative theories, about which we know that they are unitary. In this way we achieved unitarity, but the dynamics is modified from the simple two-derivative to a renormalizable dynamics with higher derivatives. Additionally the appearance of the form-factor makes the theory strongly UV convergent without the need to modify the spectrum or introducing ghost instabilities. One thing maybe needs to be explained better here. Despite that in the UV regime we recover polynomial higher derivative theory, the analysis of tree-level spectrum still gives us a unitary theory without ghosts. The explanation of this apparent paradox comes from the fact that renormalizability is the behaviour of the theory in the very UV limit, while unitarity is influenced by the behaviour at any energy scale. In a sense 
 unitarity is a global notion on the real line of energy scales, while renormalizability is a local one defined only at $k=+\infty$. Naively from the polynomial UV behavior we would derive the existence of additional poles at some finite energy scales, but exactly in this regime the UV polynomial theory is not valid. The correct behaviour at these scales is given by the full non-local theory and such is unitary. In our theory unitarity is secured at all energy scales because we use the interpolating function $V^{-1}(z)$ given by an entire function without any zeros on the complex plane.

\subsection{
Quantum divergences} 
In the high energy regime (in the UV), 
the propagator for both theories in momentum space 
schematically scales as 
\be
\mathcal{O}^{-1}(k) \sim k^{-(2 \gamma +2 \mathrm{N} +4)} \,. 
\label{OV} 
\ee
The vertices of the theory can be collected in different sets, that may involve 
or not the entire function $\exp H(z)$. 
However, to find a bound on the quantum divergences it is sufficient to concentrate on
the leading operators in the UV regime. 
These operators scale as the propagator, they can not have higher power of momentum $k$ in scaling, in order not to break the renormalizability of the theory. The consideration of them gives the following 
upper bound on the superficial degree of divergence of any graph  \cite{modesto,A},
\be
\omega(G)\leq DL+(V-I)(2 \gamma +2 \mathrm{N} +4)\,.
\ee
This bound holds in any spacetime of even or odd dimension. After plugging the definition of capital $\rm N$,  in even dimensions we get the following simplification:
\be 
&&
\omega(G) \leq  D - 2 \gamma  (L - 1)    \, , \,\,\,\,
\label{even}
\ee
We comment on the situation in odd dimensions below.
Above we used the topological relation between the numbers of vertices $V$, internal lines $I$ and 
number of loops $L$ valid for any graph $G$: $I = V + L -1$. 
Thus, if $\gamma > D/2$, in the theory only 1-loop divergences survive.  
Therefore, 
the theory is one-loop super-renormalizable \cite{Krasnikov, Tombo, Efimov, Moffat3,corni1}
and only a finite number of operators of mass dimension up to $M^D$ 
has to be included in the action in even dimension to absorb all perturbative divergences. 

In gauge theory the scaling of vertices originating from kinetic terms of the type $F({\cal D}^2)^{\mathrm{N}+ \gamma+1}F$ is lower than the one seen in the inverse propagator $k^{2 \gamma +2 \mathrm{N} +4}$. This is because when computing variational derivatives with respect to the dimensionful gauge potentials (to get higher point functions) we decrease the energy dimension of the result. Hence the number of remaining partial derivatives, when we put the variational derivative on the flat connection background, must be necessarily smaller. This means that we have a smaller power of momentum, when the 3-leg (or higher leg) vertex is written in momentum space. (The maximal scaling we get for gluons' 3-vertex in the theory and it is with the exponent $2\gamma+2{\mathrm N}+3$.) In this way we can put an upper bound on the degree of divergence for higher derivative gauge theories even with a little excess. Again also in the case of higher derivative gauge theories for $\gamma>D/2$ we confidently have one-loop super-renormalizability.

In {\em odd number of dimensions} we can easily show that the theory is completely finite. If we use the dimensional regularization scheme (DIMREG) for regularizing possible logarithmic divergences \cite{A}, we quickly see that {\em there are no divergences at one loop
and the theory is automatically finite}. The reason is of dimensional nature. In odd dimension the energy dimension of possible one-loop 
counterterms needed to absorb logarithmic divergences can be only odd. However, 
at one-loop such counterterms can not be constructed in DIMREG scheme and having at our disposal only Lorentz invariant (and gauge covariant) building blocks that always have energy dimension two. By elementary building blocks we mean here curvatures (field strengths) or covariant box operators (also gauge covariant), or even number of covariant derivatives (even number is necessary here to be able to contract all indices). For details we refer the reader to original papers \cite{modesto}.

\subsection{Finite gravitational and gauge theories in even dimension}
The main reason to introduce a potential $\mathcal V$ in the action in even dimension is to make the theory finite at quantum level. We concentrate on the last set of operators as listed in (\ref{K0}) and so called killers of beta functions. It is easy to see that it is always possible to choose the non-running coefficients $s_{k,i}$ in (\ref{K0}) in such a way
to make all the beta functions to vanish. Let us expand on this point for the simpler case of
``monomial" UV asymptotic behavior of the form-factor, namely: $p_{\gamma + {\rm N} +1} (z) = z^{\gamma + {\rm N} +1}$. 
For this particular choice of the form-factor in the large $z$ limit the analysis of the previous section shows that only divergences with $D$ derivatives (like $\frac{1}{\epsilon}\mathcal{R}^{D/2}$ for the case of gravity) are generated at one loop level. We remind that for a sufficiently large choice of the $\gamma$ parameter we have perturbative divergences only at one loop. The divergences, as mentioned above, give rise to the RG running of only the dimensionless coupling constants of the theory. The kinetic terms of the theory contribute to the beta functions of these couplings in a quite non-trivial way (rational functions of frontal kinetic coefficients.) However, the last set of operators in (\ref{K0}) gives a contribution to the beta functions linear in the 
parameters $s_{k,i}$, so they can be fixed to make the total beta functions zero. In this way we achieve a theory finite in any dimension at quantum level and at all orders in the loop expansion. Let us consider as an example the situation in the familiar four-dimensional spacetime.
We need only one killer operator for a gauge theory and two killers for gravity because of the possible tensorial structures of the counterterms in $D=4$. The procedure, described below, of obtaining quantum finite theories from one-loop super-renormalizable ones is universal. It applies equally well to gravitational, gauge theories, and even to scalar and fermion theories (like the matter sector of the standard model of particle physics that is considered in detail in section three.) This is actually the reason for the title of our paper.

\paragraph{Four dimensional theory}
In $D=4$ the whole situation is simple 
 to describe. 
The highest derivative terms in the kinetic part of
the action all come from the UV limit of the form-factor and 
are of the type ${\cal R} \, \square^{\gamma}{\cal R}$ \cite{shapiro3}
or  ${ F} \, {({\cal D}^2)}^{\gamma+ 1}{ F}$. 
If $\gamma=0$, then we have only renormalizability and 
the divergences must be absorbed at every loop order. 
For super-renormalizability we need $\gamma\geq 1$: 
for $\gamma=1$ we have
3-loop super-renormalizability (so no divergences at $4$ loops and higher); for
$\gamma=2$ we have 2-loop super-renormalizability and finally starting
from $\gamma=3$ we have one-loop super-renormalizability. We choose $\gamma\geq 3$ and therefore quantum divergences can appear at most at one loop. 

Now increasing
the value of $\gamma$ does not improve the situation, however we can
ask easily for finiteness of the theory. In gravity, divergences at one-loop cause
the need for the renormalization 
and the introduction of the counterterms for $\bar{\lambda}$ (cosmological constant), $R$ (Planck constant), $R^{2}$ 
and $R_{\mu\nu}^{2}$ (two quadratic couplings).\footnote{In four dimensions another quadratic in curvature invariant ${\rm Riem}^2$ is not independent and can be always reduced to the two terms $R^{2}$, $R_{\mu\nu}^{2}$ and a Gauss-Bonnet term, which is hovewer a topological quantity. Moreover we neglect a total derivative term $\Box R$.} 
For gauge bosons we only have to renormalize $F^2$. 
Let us summarize these couplings in order
saying which operators contribute to which divergences. We will describe
the operators by giving their total number of derivatives acting on
the metric tensor (or a gauge fluctuation field) and giving the number of curvature (field strengths) tensors involved.

On the running of the cosmological constant only the operators quadratic in the curvature have impact. 
The terms with $2\gamma$ derivatives give a contribution linearly proportional to their frontal
coefficients, while terms with $2\gamma+2$ derivatives
give contributions quadratically dependent on their coefficients. 
As it will be shown elsewhere, it is easy to find such a combination
for non zero values of the coefficients of these operators to make the cosmological beta function vanish
\cite{MathStructure}. 

The running of the Planck scale parameter is simpler. There are two
contributions that are linearly proportional to frontal coefficients of
the corresponding terms. From the quadratic in curvature terms there is
one relevant type of terms with $2\gamma+2$ derivatives. 
If a more general potential $\mathcal{V}$ includes cubic operators in the curvature,
then it  
also contributes with such operators 
and again with $2\gamma+2$ derivatives. 
Therefore, it is possible to
solve one linear equation expressing the condition for the vanishing of
the beta function and to find the values of the coefficients of cubic terms. 
One cubic killer with specially chosen coefficient 
 does the job of killing this beta function for the Newton constant.

The beta functions for terms quadratic in curvatures are complicated, 
but all these contributions come from terms with $2\gamma+4$
derivatives on the metric. First, there are contributions coming from the
highest derivative terms in the kinetic part, so with $2\gamma+4$
derivatives (from the two operators of the type ${\cal R}\square^{\gamma}{\cal R}$ in the UV expansion of the form-factor). Dependence
on their coefficients is given by quite nonlinear functions. Actually
these are rational functions due to the presence of denominators related
to propagators. Second, there are also contributions quadratically dependent
on the coefficients of cubic in curvature operators (if any) in the potential.
Last there are contributions coming from operators quartic in curvature.
These terms contribute in a linearly dependent way in their coefficients. 
The full system of equations for the two conditions of vanishing of beta functions 
is fairly easy to solve for the coefficients of quartic
operators, even in the absence of cubic operators. And this is exactly,
what we recently did in \cite{modestoLeslaw}.

For the gauge theory (\ref{gauge}) we have just to repeat the analysis done for the 
operators quadratic in curvature for gravity. We may concentrate on the introduction of one operator quartic in the field strength 
${F}^2 {({\cal{D}}^2)}^{\gamma-1} { F}^2$ in order to kill the beta function of the single YM coupling (this coupling stands in front of the two-derivative kinetic term $F^2$.)
 
Let us here summarize what we need to make finite a super-renormalizable theory of gravity 
in $D=4$. First the coefficients in front of all four
terms of the type ${\cal R} \, \square^{\gamma-1}{\cal R}$ and ${\cal R} \, \square^{\gamma-2}{\cal R}$
must be set to zero to avoid running of the cosmological
constant. Due to the strong convergence property (\ref{Vlimit1}) this corresponds to the minimal choice of the asymptotically polynomial form-factor,
namely $p(z) = z^{\gamma+1}+O(z^{\gamma-2})$. 
Then it is optional or to put to zero all frontal coefficients
for terms cubic in curvature and with $2\gamma+2$ derivatives, either
to solve the linear equation for vanishing of the beta function for the 
Newton constant. This last option would express one coefficient
in terms of a linear dependence on all the others. 
By adjusting the parameters
of the theory in the potential to satisfy this choice, we get rid of perturbative
running of the Newton constant.
In order to kill the running of coupling constants in front of the operators quadratic
in curvature we also face with multiple choices. The minimal one is to
set to zero all cubic operators and to invoke only two terms that
we call killers of the beta functions and are 
quartic in curvature.
The richer option is to take into account
all possible cubic and quartic operators at this order of $2\gamma+4$ 
derivatives of the metric. 
Now we may use two linear relations to make the two parameters
for the quartic operators dependent on all other parameters of terms quadratic,
cubic and quartic in curvature. (It is not known, if the same can be
achieved with only cubic operators.) By adjusting the two parameters
in the theory to satisfy this choice, we get rid of perturbative
running of the two coupling constants for the quadratic operators,
$
\beta_{R_{\mu\nu}^2} \, \,\, {\rm and} \,\, \,
\beta_{R^2} 
$ are therefore zero.

Besides this there are
no other conditions to be imposed on those terms, if we demand perturbative
finiteness of the theory, provided that the conditions for one-loop
super-renormalizability are satisfied. We decided to work here with the minimal choice.
Therefore, the conditions for finiteness reduce only to a relation between the front coefficients for the 
killers $s_1$ and $s_2$ 
and two other constants $c_{1}$ and $c_{2}$ 
that have to be determined 
calculating the contributions to the beta functions 
from the terms quadratic in curvature. 
We compute them using Barvinsky-Vilkovisky (BV) techniques 
\cite{GBV, Shapirobook, modestoLeslaw}. We neglect here all the details of this very powerful method for generally covariant computation of one-loop effective action in 4-dimensional renormalizable quantum field theories. We only advert, that the beta functions of our theory appear as the coefficients in front of the operators in the divergent part of quantum effective action.
It is crucial that all the operators 
involved in the divergent contributions to the effective action do not take infinite renormalizations and therefore no running coupling constants appear in the expressions for the beta functions. Since we assume $\gamma>2$ all the operators of dimension up to $M^4$ in the bare action do not give contributions to the divergent part of the one-loop effective action.  

The situation with gauge theory is even simpler. From the same reason - to avoid running of vacuum energy - we choose a polynomial $p(z)$ to be a monomial. In this minimal setup the monomial in UV gives precisely the highest derivative term of the form $F({\cal D}^2)^{\gamma+1} F$. There is only one possible way how to take trace over group indices here, and terms with derivatives can be reduced to ones with gauge covariant boxes only by exploiting Bianchi identities in gauge theory. The contribution to the running of the YM coupling from this quadratic term is actually a dimensionless constant (independent of the frontal coefficient of the highest derivative term), which again has to be determined by a computation using BV technology. This number can be cancelled by a contribution coming from a quartic gauge killer of the form 
\be
s_g  F^2({\cal D}^2)^{\gamma-1} F^2 \,\,\,\, (\mbox{GAUGE -- THEORY KILLER})
\ee
(here there are many possibilities of taking trace(s).) This contribution is linear in the parameter $s_g$ and hence the latter one can be adjusted to make the total beta function vanish.

Below we present an explicit example of a finite theory for gravitational interactions. The minimal choice for a finite and unitary theory of quantum gravity in four
dimension may consist of terms with $\gamma=3$ in the kinetic
part. 
For simplicity we may have only
two killers and no cubic in curvature operators. 
Given the entire function $H(z)$,
\be
&& \hspace{-1cm} H(z) = \frac{1}{2} \left[ \Gamma \left(0,  p(z)^2 \right)+\gamma_E  + \log \left( p(z)^2 \right) \right]  ,
\label{MinimalH}
\ee
the simplest Lagrangian 
may be the following,
\be
&&
 \hspace{-1.4cm} 
 \mathcal{L}_{\rm fin,\,gr} = - 2 \kappa_4^{-2}   \Big[ R  
+ R_{\mu \nu} \,  \frac{ e^{H(-\Box_{\Lambda})} -1}{\Box}   R^{\mu \nu} 
- \frac{1}{2} R \,  \frac{ e^{H(-\Box_{\Lambda})} -1}{\Box}   R
+s_{1}R^{2}\square R^{2}+s_{2}R_{\mu\nu}R^{\mu\nu}\square R_{\rho\sigma}R^{\rho\sigma} \Big]  , 
\label{Minimal}
\ee
where 
\be
&& 
p(z) = z^{\gamma+1}=z^4 \, , \,\,\,\, (\gamma =3) \, , \nonumber \\
&&
s_{1}=-\frac{2\pi^2}{3}\omega_{2}(c_{1}+c_{2}) \,, \,\,\,\,  s_{2}=8\pi^2\omega_{2}c_{2}
\,\,\, \,\, \mbox{and} \,\,\,\,\, 
\omega_2  = \frac{e^{\gamma_E/2}}{\Lambda^{2 \gamma+2}}= \frac{e^{\gamma_E/2}}{\Lambda^8}.
\ee
%
A more general Lagrangian can have a bunch of other terms (but still
finiteness of the theory can be obtained exactly in the same way):
\be
&& \hspace{-0.5cm}\mathcal{L}_{\rm fin,\,gr} = - 2 \kappa_4^{-2}   \Big[ R  - \frac{\bar{\lambda}}{2 \kappa_4^{-2}} 
+ G_{\mu \nu} \,  \frac{ e^{H(-\Box_{\Lambda})} -1}{\Box}   R^{\mu \nu} 
+s_{1}R^{2}\square R^{2}
+s_{2}R_{\mu\nu}R^{\mu\nu}\square R_{\rho\sigma}R^{\rho\sigma} \Big]
\nonumber \\
&& \hspace{1.7cm}
+ \sum_i c_{i}^{(3)}\left({\cal R}^{3}\right)_{i}+ \sum_i c_{i}^{(4)}\left({\cal R}^{4}\right)_{i}+ 
\sum_i c_{i}^{(5)}\left({\cal R}^{5}\right)_{i}. 
\ee
In the first line we have added a cosmological constant term $\bar{\lambda}$. Note that in the last line there are no covariant derivatives, 
and that $c_{i}^{(3)}$, $c_{i}^{(4)}$ and $c_{i}^{(5)}$ are some constant coefficients.
In the case of gauge theory the action of the finite quantum theory may take the following compact form (for the choice $\gamma=3$): 
\be
\mathcal{L}_{\rm fin,\,YM} = -\frac{1}{4g^2}\Big[ \underbrace{F_{\mu\nu}e^{H(-{\cal D}_\Lambda^2)}
F^{\mu\nu} + s_g F^2({\cal D}^2)^{2} F^2}_{\mbox{minimal finite theory}} + \sum_i \sum_{j>2}^6 \sum_{k=0,\,k<4}^{6-j} c^{(j,k)}_i \left(({\cal D}^2)^k F^j\right)_i\Big]\,.
\ee
Again $c^{(j,k)}_i$ are some constant coefficients. The last terms in the last two formulas have been written in a compact index-less notation.

\paragraph{An elegant \& finite gravitational theory}

We recently discovered another class of weakly non-local theories entirely kinematical (where only 
weakly non-local operators quadratic in curvature appear) without local or non-local potentials cubic in curvature or higher. The simplest such four dimensional theory reads as 
(this is partially motivated by the recently discovered work \cite{Kuzm})
\be
&& \hspace{-1cm}
\mathcal{L} = - 2 \kappa^{-2}_4 
\sqrt{ - g} \left[  R + \gamma_0 R f(\Box) R + \gamma_2 R_{\mu\nu} f(\Box) R^{\mu\nu}
+ \gamma_4 R_{\mu\nu\rho\sigma}  f(\Box) R^{\mu\nu\rho\sigma}
+ \gamma_W C_{\mu\nu\rho\sigma}  f(\Box) C^{\mu\nu\rho\sigma}
\right] \,,  \nonumber \\
&&\hspace{-1cm}
 f(\Box) = \frac{e^{H(-\Box_\Lambda)} - 1}{\Box} \, . 
\ee
We have added above other kinetic terms made of the Riemann and Weyl tensors respectively. Moreover $\gamma_0, \gamma_2, \gamma_4, \gamma_W$ are some parameters. 
We get quite easily one-loop super-renormalizability 
only requiring $\gamma\geq 3$ in the UV polynomial as in (\ref{Vlimit1}). We have the freedom in choosing four couplings $\gamma_0, \gamma_2, \gamma_4, \gamma_W$. One condition must be imposed on them to have aunitary theory (with multiplicative modification of the propagator from Einstein's theory). The two next conditions are requirements for vanishing of the two beta functions. We are left with 
the freedom of an overall multiplicative factor for the quadratic in curvature part of the action. By naive counting it seems possible to satisfy all three conditions and to finally get the finite theory. 

There is a hope that by choosing well adjusted values of the non-running coefficients $\gamma_0, \gamma_2, \gamma_4, \gamma_W$ we may achieve finiteness of this theory. At the moment we can not definitely assert this issue because the beta functions for the couplings in front of the operators $R^2$ and $R_{\mu\nu}^2$ are rational functions in all four parameters with numerators, which are quadratic in them. Here instead of the roles played by two killers we have two additional kinetic operators - they do not only contribute to the propagator (which can be read entirely from the reduced action with the only two simplest kinetic quadratic terms), but also to vertices of the theory and generating in a sense hidden killers. The above expectation for a finite theory can be even strengthened, when we include a generalized Gauss-Bonnet term. We construct it by putting in the middle the form-factor $f(\Box)$ in each of the term resulting after decomposition of Gauss-Bonnet term into expressions explicitly quadratic in curvature, namely
\be 
{\rm GB}_{f} = 
R_{\mu \nu \alpha \beta } \, f(\Box) R^{\mu \nu \alpha \beta}    
\label{propertyh4} 
- 4  R_{\mu \nu} \, f( \Box) R^{\mu \nu} 
+ R \, f(\Box) R \, .
\ee
Such term does not completely influence the graviton propagator around flat spacetime. It can give rise only to vertices of the theory, which here behave like killers. We may adjust the frontal coefficient of the generalized 
GB$_f$ term in such a way to cancel one beta function, without taking the risk of trivializing the theory. This theory is elegant because we do not use added explicitly and somehow by hand killers to kill the beta functions. This theory and its generalizations, if needed, will be studied and published in a future paper.


\subsection{Scale invariance \& Conformal symmetry}
We want to comment here on the issue of scale invariance vs. conformal invariance in our candidate finite theory. The theory is unitary and scale-invariant at quantum level because all the beta functions are identically zero. This has already implications for the Ward identities of the scaling symmetry involving the trace of the energy tensor as read from the full effective action. However, we recall that the theory we started with is nor scale neither conformally invariant at classical level. For one of the first time we meet a situation, in which, due to quantum effects,
scale-invariance is restored. Thanks to some theorems in $D=4$ also conformal symmetry 
should be hidden here \cite{Polchinski}. As it is known in QFT, vanishing of all beta functions means that scale invariance is realized as a symmetry on the quantum level. Therefore, we expect even more, namely the full effective action will enjoy full conformal symmetry understood on the quantum level through conformal Ward identities.

Of course in the full effective action we have also non-universal finite terms appearing at every loop order. But thanks to the emerging conformal symmetry we may find a unique effective action with completely specified form of these non-divergent terms. Only such effective action will be conformally invariant on both classical and quantum level. In field theory  we conclude solely from beta functions about the presence or spontaneous violation of scale invariance. Finite terms must respect these findings about scaling symmetry, when the theory is written in a conformally invariant fashion. This may allow us in the future to find the full form of the effective actions (with explicit form of finite terms.) Right now we know about this effective action that the divergent part is not there (the same for the counterterm Lagrangian) and that the bare action is given by the classical action (\ref{gravity}). We think that it is also possible to find the missing form of the finite terms in the effective action by looking for classically conformally invariant gravitational actions in finite dimensional spacetimes (for the case of the minimal finite theory (\ref{Minimal}) we would have to consider $D=10$).

The phenomenon of emerging conformal symmetry on the quantum level can be understand also 
in a different way. We can assume that contributions from terms classically violating conformal symmetry in the classical action are exactly cancelled by loop contributions representing the quantum violation of scale invariance. In other words we can say that classical Weyl anomaly compensates precisely the quantum Weyl anomaly and in the result the full quantum effective action (with tree and all loop levels) is without such conformal anomaly at all. 
Moreover, in the finite theory on the quantum level we expect the trace of the energy tensor derived from the effective action to be identically zero \cite{DeserConformal}, 
\be
T^\mu_{\mu} := g^{\mu \nu} \frac{\delta \Gamma_{\rm eff}}{\delta g^{\mu \nu}} \equiv 0.
\ee
Yet another interpretation is that the finite theory sits exactly at the non-trivial fixed point (FP) of RG flow. Therefore it can be used to describe the physics near the FP in conformal perturbation calculus. Remarkably one of such finite gravitational theories gives the UV FP action for the quantum gravitational theory, whose infrared (IR) limit is Einstein's classical gravity.
%
\subsection{Bound states in super-renormalizable theories}
It is worth of mentioning a very exciting perturbative high energy picture proposed in 
\cite{Tombo}.
The class of theories proposed in this paper 
can be expected to have a spectrum of perturbative (because we are dealing with super-renromalizable and therefore asymptotically free theories) bound states, thus creating their ``own matter". These may be bound states of gauge bosons or gravitons, as well as of externally introduced matter.
This is supported by the behaviour of the propagator 
at length scales of order $1/\Lambda$ or smaller. Indeed in the deep UV regime the inverse bare gauge boson or graviton propagator is given by a polynomial of degree $2 \gamma+D$ in momentum, corresponding to a tree-level confining 
potential between particles: $ V( r ) \sim -{\rm c} + {\rm c}' r^2$, with c and c$^\prime$ positive constants (it is obtained from the propagator (\ref{propagator}) for any weakly nonlocal form factor \cite{Vincent}.)  
For sufficiently large $\Lambda$ the potential can be made extremely steep. 
Therefore, systems bound by the short distance confining potential will thus be effectively permanently confined, and they will form string-like excitations with almost linearly rising spectrum (like on the leading Regge trajectory): $M^2 \propto \Lambda^2 J^{\alpha}$, ($\alpha \approx 1$, compare with \cite{Tombo}.)

Moreover for the case of a super-renormalizable theory the finite quantum logarithmic corrections to the propagator 
gives rise to an infinite number of complex conjugate poles in the so called dressed propagator. 
These extra modes have an almost equally spaced masses squared and, perhaps, correspond naturally 
to an infinite tower of massive states - the same like always seen in string theory \cite{Barnaby, shapirocomplex}. 

\section{Coupling of matter}
In this section we expand about the need for a weakly non-local extension of the standard model of particle physics. As it is widely known the standard model is a gauge theory to which matter is coupled in two sectors: fermionic sector (of quarks and leptons) and scalar sector (of Higgs field). If using above procedure we obtain finite pure gauge theory, then the natural question arises, whether this nice property will survive after coupling to matter. The same question can be asked on the level of coupling the matter sector (understood as non-gravitational) to the finite gravitational theory. We will here mainly address the second version of this question, but the same answers would apply also to the case of pure gauge theory. Following the results in \cite{shapiro3} it turns out that the gravitational sector proposed in this paper does not give any divergent contribution to the beta functions in matter sector provided that values of the integer $\gamma$ satisfy the inequality $\gamma >1$. Therefore, 
the standard model beta functions remain untouched after coupling to gravity. (This also means that standard model remains a renormalizable theory on curved spacetime as it was on Minkowski spacetime.) Conversely, the interactions of matter fields of the local standard model change the gravitational beta functions at any order in the loop expansion invalidating the super-renormalizability property of pure gravity. The full theory is still renormalizable, while the matter 
contribution to the gravitational beta functions turns out to be exactly the same like evaluated for semiclassical gravity (gravitons can only be on external lines of Feynman diagrams). This phenomenon can be understood in the following way. If the two quantum sectors are coupled to each other via renormalizable interactions, then the renormalizability properties of the full theory are the minimal properties out of all that are present in the two sectors. This explains, why if the theory for matter is only renormalizable, then coupling to the super-renormalizable gravity results in only a renormalizable full theory.
The divergent contribution to the gravitational part of one-loop effective action for a general theory with $N_s$ scalars, $N_f$ spinors and $N_v$ vectors reads as follows \cite{Shapirobook},
\be
\Gamma^{(1)}_{\rm div} = \frac{1}{\epsilon}  \frac{1}{16 \pi^2} \int \! d^4 x \sqrt{|g|} \left[ \beta_{\bar{\lambda} } + \beta_R \, R + \beta_{C^2} C_{\mu\nu\rho\sigma}^2 + \beta_{R^2} R^2 \right]  ,
\ee
where $C_{\mu\nu\rho\sigma}$ denotes the Weyl tensor and the beta functions are:
\be
\hspace{0.2cm}
 \beta_{\bar{\lambda}} = \frac{1}{2} m_s^4 - 4 m_f^4 \, , \,\,\,\, 
 \beta_R = N_s m_s^2 \left( \xi - \frac{1}{6} \right) + \frac{2}{3} N_f^2 m_f^2 \, , \,\,\,\, 
\beta_{C^2} = \frac{N_s}{120} + \frac{N_f}{20} + \frac{N_v}{10} \, , \,\,\,\, 
\beta_{R^2} = \frac{1}{2} N_s \left( \xi - \frac{1}{6} \right)^2 \! \! . \nonumber
\ee
In order to derive these beta functions it was assumed that vectors and spinors are coupled minimally to gravity, while scalars were also allowed to have a non-minimal coupling of the form $\xi R \phi^2$. Scalars and fermions were taken to be massive with masses $m_s$ and $m_f$ respectively, while vectors were massless gauge fields. Moreover all these couplings are renormalizable (in the framework of two-derivative standard model and quadratic gravity). These results of semiclassical computation strongly suggest that a consistent theory of quantum gravity in $D=4$ must be formulated as a theory with higher derivative actions (at least with four derivatives) \cite{shapiro4}.

An higher derivative extension of the standard model of particle physics makes it possible to restore 
super-renormalizability for all fundamental interactions because 
all the kinetic and interaction operators in the action for matter, gravity and gauge bosons have the same scaling in the ultraviolet regime. To comply with unitarity, higher derivatives must be introduced by a weakly non-local form-factor, which has UV asymptotic polynomial behaviour. The function $H(z)$ interpolating between low energy and high energy physics must be given by some entire function. For precisely the same scaling of propagators for all quantum modes at any energy scales we are required to use only one form-factor, the same for all particles.

The weakly non-local super-renormalizable extension of the standard model leaves  open the following interesting possibilities:
\begin{enumerate}
\renewcommand{\theenumi}{\alph{enumi}.}
\renewcommand{\labelenumi}{\theenumi}
\item
The killer operators can be used 
to construct a
completely scale-invariant standard model of particle physics. As we elucidated in the previous section this finite theory will have full conformal symmetry realized on the quantum level. Actually this will be a conformal field theory of standard model interactions. Then coupling to finite gravitational theory will be easy as well.
\item 
In the coupled theory we have now more room to accommodate unification for all the four fundamental interactions. All the beta functions are independent on the running coupling constants and the free parameters in front of the killers 
can be tuned in such a way, that all the coupling parameters meet at a unification scale (at one point) without invoking supersymmetry. This signifies that in our theory unification of all interactions is easily achievable. We can also restrict a bit and demand unification of non-gravitational couplings only, this is possible too. 
\item 
 Moreover we can solve the old problems of standard model, when extended to very high energy. We easily get asymptotic freedom in UV for all interactions, with avoidance of Landau pole for electric charge coupling.
\end{enumerate}

Non-local extension of the standard model seems to be a viable theoretical possibility (even if this is not motivated by a need to consistently coupling to quantum gravity). But one question comes immediately after this step. Can we find a compatibility of predictions of our theory with that one from renormalizable standard model with local two-derivatives actions? We think that the answer to this issue is positive because even in pure QFT framework (without coupling to gravity) SM must be seen only as an effective low-energy theory with the smallest possible number of: derivatives in kinetic terms, light fields and renormalizable interaction terms. Only UV completion of SM can solve its old problems, which are the main problems of particle physics nowadays. The great experimental success of standard model is in modern view related to the fact, that this UV completion happens at very high energy scale (compared to SM scale of few hundred GeV). 

Still the effect of this completion in UV can be seen at low energies as the presence of higher derivative non-renormalizable operators in effective field theory, which is beyond SM. We can make a very naive estimation of number of derivatives in these operators, when they appear in effective action with order one frontal coefficients. This will estimate the value of $\gamma$ parameter, which is not in conflict with experimental observations of particle physics. Let's assume that some sectors of renormalizable SM (we mean electrodynamics) are in full agreement with experiments up to $L$ loop level. 
This would mean, that in the effective field theory this effect could be mimicked by higher derivative operator on a tree-level with 
$2L+2$ 
derivatives (assuming new physics at TeV scale), so with $\gamma+1=L$ in the case of gauge theory (\ref{gauge}). (In doing this estimate we used the fact, that at the level of $L$ loops we have to make $L$ four-dimensional integrals over loop momenta and that the scaling of higher derivative operator is the same for propagator and vertices derived from it. Moreover we use the information that all bosonic propagators in the SM are like $k^{-2}$ in momentum space.) For the case of $L=5$ (stringest tests of QED) we get the bound $\gamma>4$.

Another issue is related to cancellation of all types of gauge anomalies in the standard model of particle physics. This may be thought of as a pleasant accident, which may happen only in local theories with precisely tuned particle content of the model. Along with other accidental symmetries (like baryon, lepton number conservation) this is a feature that will not be preserved when we move towards higher energies because typically new heavy modes will appear there. Conversely as it was shown recently by Anselmi \cite{anselmi} the cancellation of all dangerous gauge anomalies continues to happen also in weakly non-local gauge theories.

\subsection{Finite Non-local Standard Model of particle physics}
Below we will give some details on how to make the standard model a finite theory at the  quantum level and when coupled to non-local finite gravity. For the case of the standard model of particle physics coupled to gravity the gauge group is 
${\cal G}_{\rm SM+gr} =GL(4) \times SU(3)_{\rm strong} \times SU(2)_{\rm weak} \times U(1)_Y$. We choose a non-local form-factor in the form (\ref{Vlimit1}) with UV monomial $p(z)=z^{\gamma+1}$. It is then sufficient to introduce up to $3+2$ killers to make the gravity-gauge sector of the theory finite, namely we introduce the killer operators 
\be
R^2 \Box^{\gamma -2} R^2 \,\,\,\,\,  {\rm and} \,\,\,\,\,    R_{\mu \nu}^2 \Box^{\gamma -2} R_{\mu\nu}^2
\label{gravity2}
\ee
to make vanish the beta functions for the counterterms $R^2$ and $R_{\mu\nu}^2$, which absorb all logarithmic divergences in $D=4$ in the gravitational sector.
We use other three gauge killers (one for each gauge group), with the form
\be
F_{\mu \nu}^2 {({\cal D}^2)}^{\gamma-1 } F_{\mu\nu}^2 \, ,
\label{gauge2}
\ee
to make vanish the beta functions for each of the operators $F^2$ (for each gauge coupling constant.) We end up with a completely scale-invariant theory for all fundamental interactions. Besides gauge and gravity sectors in Nature we have also low spin matter. We will give details of non-local extension of this sector in the next subsection. Right now we only comment on the number and form of possible killers in this sector. In full generality two more killers may be necessary in order to fully remove the divergences here. They are related to the renormalization of Yukawa coupling $h$ and to the Higgs boson 
$\Phi^4$ self-interaction coupling constant $\lambda$. The tentative form of these killers may be
\be
(\bar{\psi} \Phi \psi) \Box^{\gamma} (\Phi^\dagger \Phi) \quad {\rm and} \quad 
(\Phi^\dagger \Phi)^2 \Box^{\gamma} (\Phi^\dagger \Phi)\,. \ee 
We notice that the box operators in the formula above are standard covariant d'Alambertians because the expressions on both sides are invariant with respect to the gauge group for the SM. We avoid running of the mass parameter $\mu^2$ by choosing a monomial behaviour in the UV regime. 

\subsection{Grand Unified Theory}
As extensively explained in the previous section, the extra ``killer operators" can be used to make the theory finite by a proper fine tuning of their frontal coefficients.
However, we can also pick out such coefficients to tune the beta functions for the gauge couplings and Newton constant in such a way that they meet in a single point at high energy. 
For the case of the gauge group ${\cal G}_{\rm SM+gr}$ we need up to $3+2$ 
quartic killers. We can still use the purely gravitational killers (\ref{gravity2}) to make vanish the beta functions for quadratic in gravitational curvature counterterms $R^2$ and $R_{\mu\nu}^2$. On the other hand, we want to keep $G_N$ running, so we will not choose monomial behavior of the form-factor in the UV regime. The other three gauge killers (\ref{gauge2}) are used to adjust the beta functions for the gauge coupling constants. We require them to meet $G_N/G_{N,0}$\footnote{$G_{N,0}$ is a classical value of the gravitational constant, that is in IR limit.} - the dimensionless gravitational coupling constant, all together at one energy scale. This is the scale of grand unification of all interactions. For the monomial case there is no running of the Newton constant, but if we consider a completely general form-factor with trinomial limit for large $z$, 
$p(z)=a_{\gamma+1} z^{\gamma+1} + a_{\gamma} z^{\gamma}+a_{\gamma-1} z^{\gamma-1}+O(z^{\gamma-2})$, then also $G_N$ and the cosmological constant $\bar{\lambda}$ run. Preliminary analysis of the purely gravitational case shows that a consistent relation between the above three coefficients 
$a_{i}$ exists to make $\bar{\lambda}$ scale-invariant. (We remind that there are no killers that are able to kill 
the one-loop running of $\bar{\lambda}$.) We expect the same to be true also in presence of gauge bosons and matter because the gravitational contribution to the cosmological constant beta function is linear in $a_{\gamma -1}$. In the one-loop super-renormalizable theory the RG running of couplings has exactly logarithmic character because the beta functions depend only on constant parameters in front of the terms with higher derivatives (that are not subject to infinite renormalization in $D=4$). Moreover, here we take as the definition of running the flow given by beta functions as read exclusively from the logarithmically divergent part of the effective action. We neglect all other non-universal effects on running (like finite renormalization of couplings or finite subleading corrections related to threshold phenomena for massive fields.)

Finally, other contributions to the beta functions come from the 
Fermi and Higgs sectors of the standard model, but they are without any possibility to essentially change the analysis performed in the previous paragraph. In order to complete the story with non-local standard model, we present below the weakly non-local action for a fermion or a scalar (like the Higgs field $\Phi$), 
\be
\hspace{-0.5cm}
 \sum_{a}^{N_f}\bar \psi_a \, i \slashed{\cal D}_{a}  e^{H(-\slashed{\cal D}^{2}_{a,\Lambda})}
\, \psi_a
+ ({\cal D}_{\mu} \Phi)^\dagger e^{H({-\cal D}_{\Lambda}^2)} ({\cal D}^{\mu} \Phi) 
- \mu^2 \Phi^\dagger  e^{H({-\cal D}_{\Lambda}^2)}   \Phi - \lambda (\Phi^\dagger \Phi)^2. 
\ee
As we see one universal form-factor appears in all kinetic terms for all fields (and also in mass terms for initially massive fields). In this non-local model at low energy we still have the spontaneous breaking of electroweak symmetry because our theory in the IR limit coincides precisely with the local two-derivatives standard model. The only condition here is that the scale of non-locality $\Lambda$ must be bigger than the scale of symmetry breaking.

Last remark is about gauge+gravity unification in a weakly non-local field theory. To claim unification of all fundamental interactions we need only one entire function $e^{H(z)}$ to appear in the action for all the fundamental forces. Then we also have to introduce five non-zero frontal coefficients for the killers, that we can fix requiring specific behavior of the beta functions. However, the values of these killers coefficients are specified entirely in terms of the form-factor without the need to introduce any extra free parameter. Therefore, at the unification scale, we remain with one entire function $e^{H(z)}$ and one coupling constant. 

Other relevant readings about matter coupling to gravity and unification in four or extra dimensions are 
\cite{unification1, unification2,unification3}.

\section{Conclusions}
We have advanced a general class of gauge and gravitational theories compatible with quantum super-renormalizability or finiteness together with unitarity. Therefore these theories are fully consistent with quantum mechanics. The theories are defined by equations (\ref{gauge}), (\ref{gravity}) and (\ref{K0}) together with the form-factor (\ref{Vlimit1}) in a multidimensional spacetime and by (\ref{Minimal}) in $D=4$. The key point about these theories is that they are defined by weakly non-local Lagrangians.
The actions consist of a non-polynomial kinetic term (with asymptotic polynomial behavior) and a local potential of curvature $O({\cal R}^3)$, $O({ F}^3)$. Unitarity is secured because the form-factor is required to be an entire function. An explicit example of such interpolating function was given in (\ref{Vlimit1}).
For sufficiently big values of the $\gamma$ parameter, 
quantum divergences only occur at one loop and the theory is 
super-renormalizable in any dimension. Moreover, the theories in odd dimensions are automatically finite on the quantum level. In spacetimes of even dimensionality, if we make a specific choice for a restricted number of parameters in the potential, 
all the beta functions can be made to vanish and the theory turns out to be completely finite. 
On the quantum level scale invariance  
appears and may be enhanced to full conformal symmetry. The finite theory was explicitly constructed for the case of gravitational theory in dimension four, but can be easily generalized to any dimension. The case of gauge theory has been done in perfect analogy to gravity, therefore proving the universality of our method. Moreover, these results have been exported to theories that couple to matter with implications for a quantum finite standard model of particle physics and/or unification of all fundamental interactions. In this way:

{\em we have explicitly shown how to construct a finite theory of quantum gravity unified with gauge interactions and matter}.
%

\end{document}